\documentclass{aa}

\usepackage[varg]{txfonts}
\usepackage{natbib}
\usepackage[english]{babel}
\usepackage[utf8]{inputenc}
\usepackage[T1]{fontenc}
\usepackage{xspace}
\usepackage{hyperref}
\usepackage{xcolor}
\usepackage{nicefrac}

\usepackage{siunitx}
\hypersetup{%
  colorlinks=true,
  linkcolor=blue,
  urlcolor=blue,
  citecolor=blue,
  bookmarksnumbered=true,
  bookmarksopenlevel=1,
  }

\newcommand\df{\mathrm{d}} 
\newcommand\diff[2]{\frac{\df #1}{\df #2}} 
\newcommand\Tcor{\ensuremath{T_{\mathrm{cor}}}}
\newcommand\Tcorint{\ensuremath{T_{\mathrm{cor,int}}}}
\newcommand\Tcorext{\ensuremath{T_{\mathrm{cor,ext}}}}
\newcommand\Tch{\ensuremath{T_{\mathrm{ch}}}}
\newcommand\rhoch{\ensuremath{\rho_{\mathrm{ch}}}}
\newcommand\rhochint{\ensuremath{\rho_{\mathrm{ch,int}}}}
\newcommand\rhochext{\ensuremath{\rho_{\mathrm{ch,ext}}}}

\begin{document}

\title{Cut-off of transverse waves through the solar transition region}
\titlerunning{Cut-off of transverse waves}
\authorrunning{G. Pelouze et al.}
\date{Received 23 September 2022 / Accepted 3 January 2023}

\newcommand{\orcid}[1]{}
\author{%
{Gabriel Pelouze}\inst{\ref{aff:IAS},\ref{aff:CmPA}}\orcid{0000-0002-0397-2214}
\and {Tom Van Doorsselaere}\inst{\ref{aff:CmPA}} \orcid{0000-0001-9628-4113}
\and {Konstantinos Karampelas}\inst{\ref{aff:CmPA},\ref{aff:northumbria}} \orcid{0000-0001-5507-1891}
\and {Julia M. Riedl}\inst{\ref{aff:CmPA}} \orcid{0000-0002-2327-3381}
\and {Timothy Duckenfield}\inst{\ref{aff:CmPA}} \orcid{0000-0003-3306-4978}
}

\institute{%
\label{aff:IAS}{Université Paris-Saclay, CNRS,  Institut d'astrophysique spatiale, 91405, Orsay, France}
\and \label{aff:CmPA}{Centre for mathematical Plasma Astrophysics, Department of Mathematics, KU Leuven, Celestijnenlaan 200B, 3001 Leuven, Belgium.}
\\ \email{tom.vandoorsselaere@kuleuven.be}
\and \label{aff:northumbria}{Department of Mathematics, Physics and Electrical Engineering, Northumbria University, Newcastle upon Tyne, NE1 8ST, UK}
}

\abstract%
{
Transverse oscillations are ubiquitously observed in the solar corona, both in coronal loops and open magnetic flux tubes.
Numerical simulations suggest that their dissipation could heat coronal loops, counterbalancing radiative losses.
These models rely on a continuous driver at the footpoint of the loops.
However, analytical works predict that transverse waves are subject to a cut-off in the transition region.
It is thus unclear whether they can reach the corona, and indeed heat coronal loops.
}
{
Our aims are to determine how the cut-off of kink waves affects their propagation into the corona, and to characterize the variation of the cut-off frequency with altitude.
}
{
Using 3D magnetohydrodynamic simulations, we modelled the propagation of kink waves in a magnetic flux tube, embedded in a realistic atmosphere with thermal conduction, that starts in the chromosphere and extends into the corona.
We drove kink waves at four different frequencies, and determined whether they experienced a cut-off.
We then calculated the altitude at which the waves were cut-off, and compared it to the prediction of several analytical models.
}
{
We show that kink waves indeed experience a cut-off in the transition region, and we identified the analytical model that gives the best predictions.
In addition, we show that waves with periods shorter than approximately \SI{500}{s} can still reach the corona by tunnelling through the transition region, with little to no attenuation of their amplitude.
This means that such waves can still propagate from the footpoints of loop, and result in heating in the corona.
}
{}


\renewcommand{\figureautorefname}{Fig.}
\renewcommand{\sectionautorefname}{Sect.}
\renewcommand{\subsectionautorefname}{Sect.}
\renewcommand{\equationautorefname}{Eq.}

\maketitle

\section{Introduction}
\label{sec:introduction}

Recent advances in observations and modelling have shown that magnetohydrodynamic (MHD) waves could significantly contribute to the heating of the solar corona (see review by \citealp{VanDoorsselaereEtAl2020}). In particular, transverse waves are ubiquitously observed, and they come in several kinds. The type that was first discovered are the transverse waves that are impulsively excited after a flare \citep{NakariakovEtAl1999}. However, these transverse waves are only sporadically excited and do not play an important role in the energy budget of the solar corona \citep{TerradasArregui2018}. Later on, it was discovered that the corona is filled by small-amplitude transverse waves \citep{TomczykEtAl2007, TomczykMcIntosh2009, McIntoshEtAl2011, TianEtAl2012}. These were observed in coronal loops as propagating \citep{TiwariEtAl2019} or standing waves \citep{AnfinogentovEtAl2015}. These low-amplitude transverse waves were also observed as propagating waves in open-field regions \citep{ThurgoodEtAl2014, MortonEtAl2015}. These low-amplitude waves show little-to-no decay \citep{MortonEtAl2021} and are thus named ``decayless''.

Because the flare-excited standing waves are rapidly decaying \citep{GoddardEtAl2016,NechaevaEtAl2019} due to resonant absorption \citep{GoossensEtAl2002} and non-linear Kelvin-Helmholtz instability (KHI) damping \citep{TerradasEtAl2008, AntolinEtAl2014, VanDoorsselaereEtAl2021,Arregui2021}, it is generally thought that the decayless waves must be continuously supplied with energy to counteract its strong damping. Several mechanisms for excitation have been proposed: slip-stick driving with steady flows \citep{NakariakovEtAl2016,KarampelasVanDoorsselaere2020}, vortex shedding \citep{NakariakovEtAl2009,KarampelasVanDoorsselaere2021} or footpoint driving \citep{NisticoEtAl2013,KarampelasEtAl2017} through p-modes \citep{MortonEtAl2019} or convective shuffling. The latter option of footpoint driving has had some success in generating standing mode decayless waves \citep{AfanasyevEtAl2020}, which counterbalance the non-linear damping through the KHI \citep{GuoEtAl2019multistranded} and lead to heating of loops \citep{ShiEtAl2021}.

However, for the driving of decayless waves through their footpoints, it is not well understood how the transverse waves propagate through the complicated structure of the chromosphere and transition region. The simulations of transverse-wave induced KHI heating \citep[e.g.][]{KarampelasEtAl2019stratification_resistivity_viscosity} only take into account the coronal part of the loop, that is imposing a driver at the top of the transition region. To properly model the whole loop evolution due to the wave heating, it is essential to also model the wave driver in the photosphere, and accurately capture its influence on the coronal loop dynamics. 

In plane-parallel atmospheres, the propagation of fast and slow waves has been well studied. It was found that these modes couple efficiently to Alfv\'en waves through resonant absorption \citep{HansenCally2009,CallyAndries2010,KhomenkoCally2012}. Currently, investigations are ongoing to what happens if the cross-field structuring is included into the wave propagation model \citep{CallyKhomenko2019,RiedlEtAl2019,RiedlEtAl2021loops}. Another crucial ingredient is the wave's behaviour in strong (i.e. non-WKB) stratification. It is well-known that slow waves experience a cut-off while propagating through a stratified medium \citep{BelLeroy1977}. This has been verified observationally \citep{JessEtAl2013} and numerically \citep{FelipeEtAl2018}. Still, up to now, it is unknown if a similar cut-off exists for transverse waves in structured media. For the driving of the observed decayless waves in the corona, this is a crucial property to understand.

Several analytical works predict that transverse waves are cut-off in the transition below a given frequency.
The first formula was derived by \citet{Spruit1981}:
\begin{equation}
\label{eq:wcSp81}
\omega_\mathrm{Sp81}^2 = \frac{g}{8H} \frac{1}{2\beta + 1},
\end{equation}
where $g$ is the gravity projected along the loop, $H$ the pressure scale height, and $\beta$ the ratio between the gas and magnetic pressures.
For a typical isothermal atmosphere, this corresponds to a cut-off period of \SI{700}{s} \citep{Spruit1981}.
However, \citet{LopinEtAl2014} showed that this classical cut-off is suppressed when the radial component of the magnetic field is taken into account.
\citet{LopinNagorny2017oct} later showed that transverse waves can still be cut-off, provided a non-isothermal atmosphere.
They predict the following cut-off frequency:
\begin{equation}
\label{eq:wcLN17}
\omega_\mathrm{LN17}^2 = \frac{c_{k0}^2}{4 H_0 H(z)} \left( \delta_B^2 \diff{H(z)}{z} + \frac{H^2(z)}{z^2} \right),
\end{equation}
where $z$ is the altitude, $c_{k0}$ is the kink speed at the base of atmosphere ($z = z_0$), $H$ is the pressure scale height, $H_0 = H(z_0)$, and $\delta_B^2 = \left( B_{0i}^2 - B_{0e}^2 \right) / \left( B_{0i}^2 + B_{0e}^2\right)$ is the relative difference between the magnetic field inside ($B_{0,i}$) and outside ($B_{0,e}$) the flux tube, at $z = z_0$.
Finally, an alternative formula was derived by \citet{SnowEtAl2017}:
\begin{equation}
\label{eq:wcSn17}
\omega_\mathrm{Sn17}^2 = \frac{v_A^2(z)}{4 z^2},
\end{equation}
where $z$ is the altitude, and $v_A$ is the Alfvén speed.

In this article, we modelled the propagation of kink waves in an open magnetic flux tube, embedded in a non-isothermal atmosphere. 
The atmosphere extends from the chromosphere to the corona, and includes gravitational stratification and thermal conduction (\autoref{sec:model}).
We drove kink waves at different periods, and determined whether they experienced a cut-off (\autoref{sec:res}).
We compare these results to the three analytical formulas given above in \autoref{sec:discussion}, and summarize our conclusions in \autoref{sec:conclusions}.

\section{Numerical model: magnetic flux tube through the transition region}
\label{sec:model}

\begin{figure}
\centering
\includegraphics[width=\columnwidth]{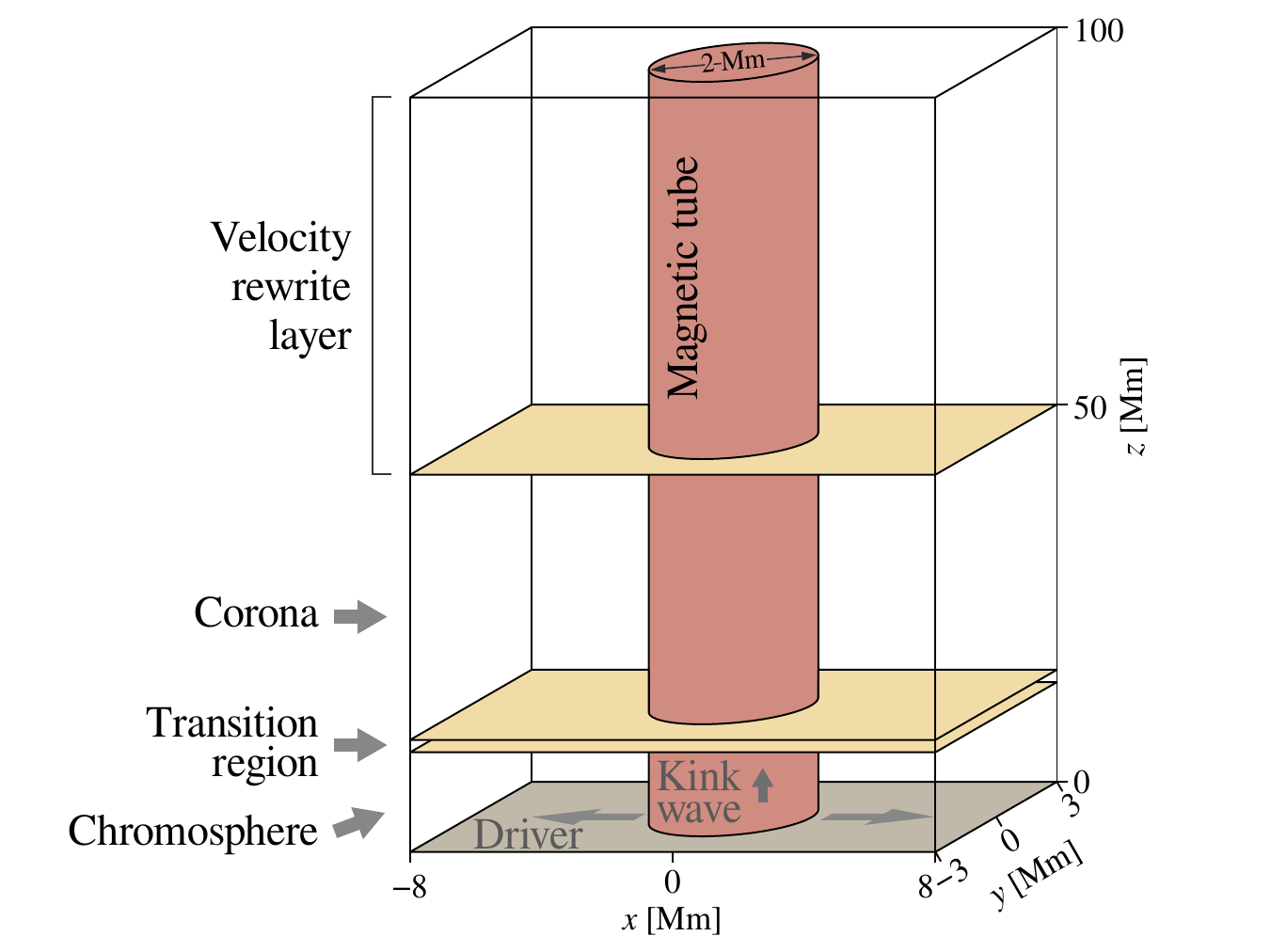}
\caption{
  Sketch of the simulation domain, showing the magnetic flux tube, the location of the kink wave driver (bottom boundary), chromosphere, transition region, corona, and velocity rewrite layer.
}
\label{fig:domain_sketch}
\end{figure}

We modelled a vertical magnetic flux tube of radius $R = \SI{1}{Mm}$ embedded in a stratified atmosphere, starting in the chromosphere (altitude $z = \SI{0}{Mm}$) and extending through the transition region ($z \approx \SI{4}{Mm}$) into the corona.
Kink waves were excited in the flux tube by applying a monoperiodic driver at the bottom of the domain ($z = \SI{0}{Mm}$).
In the upper half of the domain ($z > \SI{50}{Mm}$), we implemented a “velocity rewrite layer” to absorb the kink waves.
The driver and the velocity rewrite layer are described in \autoref{sec:model:boundary_cond}.
A sketch of the domain is shown on \autoref{fig:domain_sketch}.
We solved the 3D MHD evolution of this tube using the PLUTO code \citep{MignoneEtAl2007}, version 4.3.
This code solves the conservative MHD equations (mass continuity, momentum conservation, energy conservation, and induction equation).
We used the corner transport upwind finite volume scheme, where characteristic tracing is used for the time stepping, and a linear spatial reconstruction with a monotonized central difference limiter is performed.
The magnetic field divergence was kept small using the extended divergence cleaning method (generalized Lagrange multiplier, or GLM), and flux was computed with the linearized Roe Riemann solver.
We did not include explicit viscosity, resistivity, or cooling.
However, numerical dissipation results in higher effective viscosity and resistivity than what is expected for the solar corona, as discussed by \citet{KarampelasEtAl2019stratification_resistivity_viscosity}.
We included a modified thermal conduction, as described below.

The transition region between the chromosphere and the corona is characterized by a very sharp temperature gradient.
Resolving such gradient requires a very high resolution along the tube ($\sim \SI{1}{km}$ in the transition region).
In order to keep computational costs reasonable, we artificially broadened the transition region (thus reducing the temperature gradient).
To that end, we modified the thermal conductivity using the method developed by \citet{LinkerEtAl2001, LionelloEtAl2009, MikicEtAl2013}.
Below the cut-off temperature $T_c = \SI{2.5e5}{K}$, the parallel thermal conductivity was set to $\kappa_\parallel = C_0 T_c^{5/2}$ with $C_0 = \SI{9e-12}{Wm^{-1}K^{-7/2}}$.
Above $T_c$, $\kappa_\parallel = C_0 T^{5/2}$.
This allowed us to use a resolution of \SI{98}{km} along the tube.
This grid allows to fully resolve the broadened transition region, which has a minimum temperature scale length of \SI{1.6}{Mm} \citep[see][]{JohnstonBradshaw2019}.
The dimensions of the domain were $(L_x, L_y, L_z) = (16, 6, 100)~\si{Mm}$.
We used a uniform grid of $400 \times 150 \times 1024$ cells, with a size of $\SI{40}{km}$ in the $x$ and $y$ directions, and \SI{98}{km} in the $z$ direction.
Furthermore, we verified that the results did not change significantly when using a resolution of \SI{40}{km} in the $z$ direction. To that end, we ran a separate simulation and verified that the resulting cut-off altitude and comparison to the analytical formulas (see \autoref{sec:discussion}) were not strongly modified.
We note that such resolution is too costly in terms of compute time to be used for all simulations in this work.

The strong stratification in the transition region makes it challenging to obtain a relaxed initial state for the model.
We first initialized the domain with a field-aligned hydrostatic equilibrium (\autoref{sec:model:initial_cond}).
We then let the simulation relax in 2D for \SI{47}{ks} (\autoref{sec:model:relax_2D}).
Finally, we filled the 3D domain with this relaxed state through cylindrical symmetry, where we drove kink waves of different periods for a duration up to \SI{2.7}{ks} (\autoref{sec:model:driven_3D}).

\subsection{Boundary conditions and driver}
\label{sec:model:boundary_cond}

We first describe the boundary conditions used for the relaxation (2D) and kink wave (3D) simulations.

\paragraph{Bottom boundary} At the bottom boundary (base of the chromosphere, $z = 0$), the density and pressure were extrapolated using the hydrostatic equilibrium equation.
The magnetic field was extrapolated using the zero normal-gradient condition described by \citet[][section 2.4]{KarampelasEtAl2019stratification_resistivity_viscosity}.
For $v_z$, we either imposed a reflective boundary condition (2D relaxation, see \autoref{sec:model:relax_2D}), or imposed $v_z = 0$ (in 3D, see \autoref{sec:model:driven_3D}).
We verified that both boundary conditions give the same results in 3D simulations.
The parallel velocity components $v_x$ and $v_y$ were set to obey either a zero-gradient boundary condition (2D relaxation), or to follow a driver that excites kink waves (in 3D).
We used a monoperiodic, dipole-like, driver developed by \citet{PascoeEtAl2010} and updated by \citet{KarampelasEtAl2017}.
Inside the tube, the driver imposes:
\begin{equation}
\label{eq:driver_inside}
\left\{ v_x(x, y, t), v_y(x, y, t) \right\} = \left\{ v(t), 0 \right\},
\end{equation}
where $v(t) = v_0 \cos\left(2\pi t / P_0\right)$, with $v_0$ the driver amplitude, set to \SI{2}{\kilo\meter\per\second}.
The driver period, $P_0$, was set to different values in order to test the cut-off of kink waves.
Outside the tube, the driver imposes:
\begin{equation}
\label{eq:driver_outside}
\left\{ v_x(x, y, t), v_y(x, y, t) \right\} =
v(t) R^2
\frac{
  \left\{
  \left(x - x_0(t)\right)^2 - y^2,
  2\left(x - x_0(t)\right) y
  \right\}
  }{
  \left(\left(x - x_0(t)\right)^2 + y^2\right)^2
  },
\end{equation}
where $x_0(t) = v_0 P_0/(2\pi) \cdot \sin\left(2\pi t / P_0\right)$ is the centre of the tube’s footpoint at time $t$.
This driver generates a kink wave polarized in the $x$ direction.

\begin{figure}
\centering
\includegraphics[width=\columnwidth]{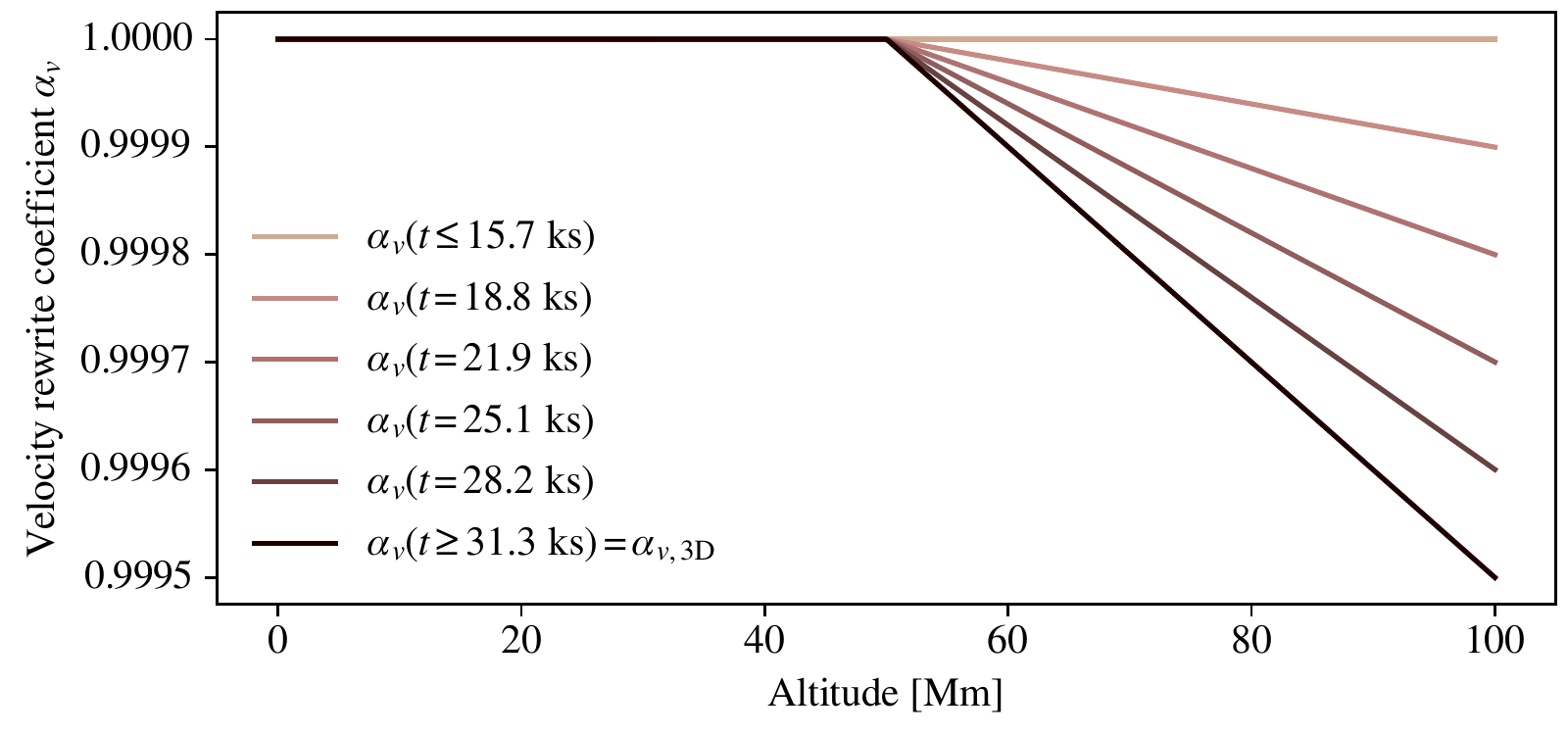}
\caption{Velocity-rewrite coefficient $\alpha_v$, applied to the velocity above \SI{50}{Mm} so that upper-propagating waves are not reflected back into the domain.
$\alpha_v$ is shown for different times of the 2D relaxation run.
The last profile ($t \geq \SI{31.3}{ks}$) is also applied in the 3D driven simulations.
}
\label{fig:vrw_av}
\end{figure}

\paragraph{Upper boundary} At the upper boundary (top of the corona, $z = \SI{100}{Mm}$),
the magnetic field was kept symmetric.
All other variables obeyed a reflective boundary condition.
In order to absorb the upwards waves excited by the driver, we artificially modified the velocity in the upper half of the domain ($z > \SI{50}{Mm}$).
At each time step, after solving the MHD equations, we decreased each component of the velocity $v_i$ by multiplying it by a quantity $\alpha_v \lesssim 1$:
\begin{equation}
\label{eq:vrw_principle}
v_i^\prime = \alpha_v(t, z) v_i.
\end{equation}
In the driven 3D simulations $\alpha_v$ was kept constant in time, and varied linearly along the loop, from 1 at $z = z_v = \SI{50}{Mm}$, to $\alpha_{v,\mathrm{min}} = \num{0.9995}$ at $z = L = \SI{100}{Mm}$:
\begin{equation}
\label{eq:vrw_av3D}
    \alpha_{v,\mathrm{3D}}(z) =
    \begin{cases}
        1 & \text{if $z \leq z_v$,} \\
        1 - \left(1 - \alpha_{v,\mathrm{min}}\right) \left(\frac{z - z_v}{L - z_v}\right) & \text{else.} \\
    \end{cases}
\end{equation}
In the 2D relaxation run, the first third of the simulation ($t_{1/3} = \SI{15.7}{ks}$) was run without modifying the velocity (i.e. $\alpha_v$ = 1).
During the second third, $\alpha_v$ was linearly ramped down in time to match the profile $\alpha_{v,3D}(z)$ described above.
Finally, the last third of the simulation was run with the constant $\alpha_{v,3D}(z)$:
\begin{equation}
\label{eq:vrw_av2D}
    \alpha_{v,\mathrm{2D}}(z, t) = 
    \begin{cases}
        1 & \text{if $t \leq t_{1/3}$,} \\
        1 - \left(1 - \alpha_{v,\mathrm{3D}}(z)\right) \left(\frac{t - t_{1/3}}{t_{1/3}}\right) & \text{if $t_{1/3} < t \leq 2 t_{1/3}$,} \\
        \alpha_{v,\mathrm{3D}}(z) & \text{else.} \\
    \end{cases}
\end{equation}
The evolution of $\alpha_v$ is shown in \autoref{fig:vrw_av}.
This “velocity rewrite layer” can successfully absorb the kink waves that are excited by the driver at the bottom of the chromosphere.
As a result, these waves are not reflected at the upper boundary, and do not propagate downwards back into the domain.
We stress that the solution obtained inside the velocity rewrite layer (i.e. above $z = \SI{50}{Mm}$) is not physical, and that this layer should be considered as a part of the upper boundary.

\paragraph{Side boundaries} At the side boundaries ($x$ and $y$ axes), all variables obeyed a zero-gradient boundary condition.
In the 2D relaxation run, we only simulated half of the tube radius ($x > 0$).
For these simulations, we imposed a reflective boundary condition on all variables at the centre of the tube ($x = 0$).

\subsection{Initial conditions: field-aligned hydrostatic equilibrium}
\label{sec:model:initial_cond}

\begin{figure*}
\centering
\includegraphics[width=0.49\textwidth]{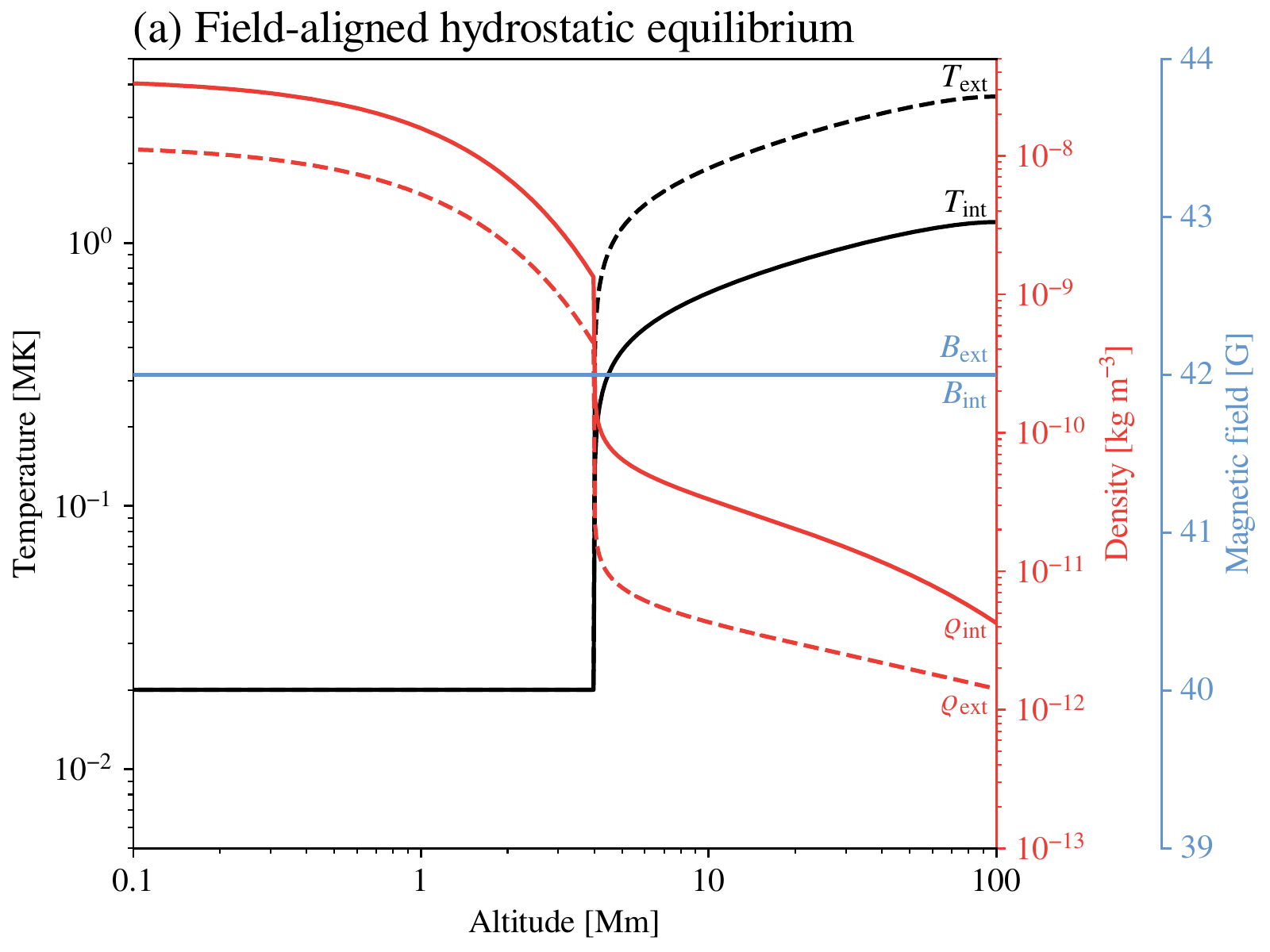}
\includegraphics[width=0.49\textwidth]{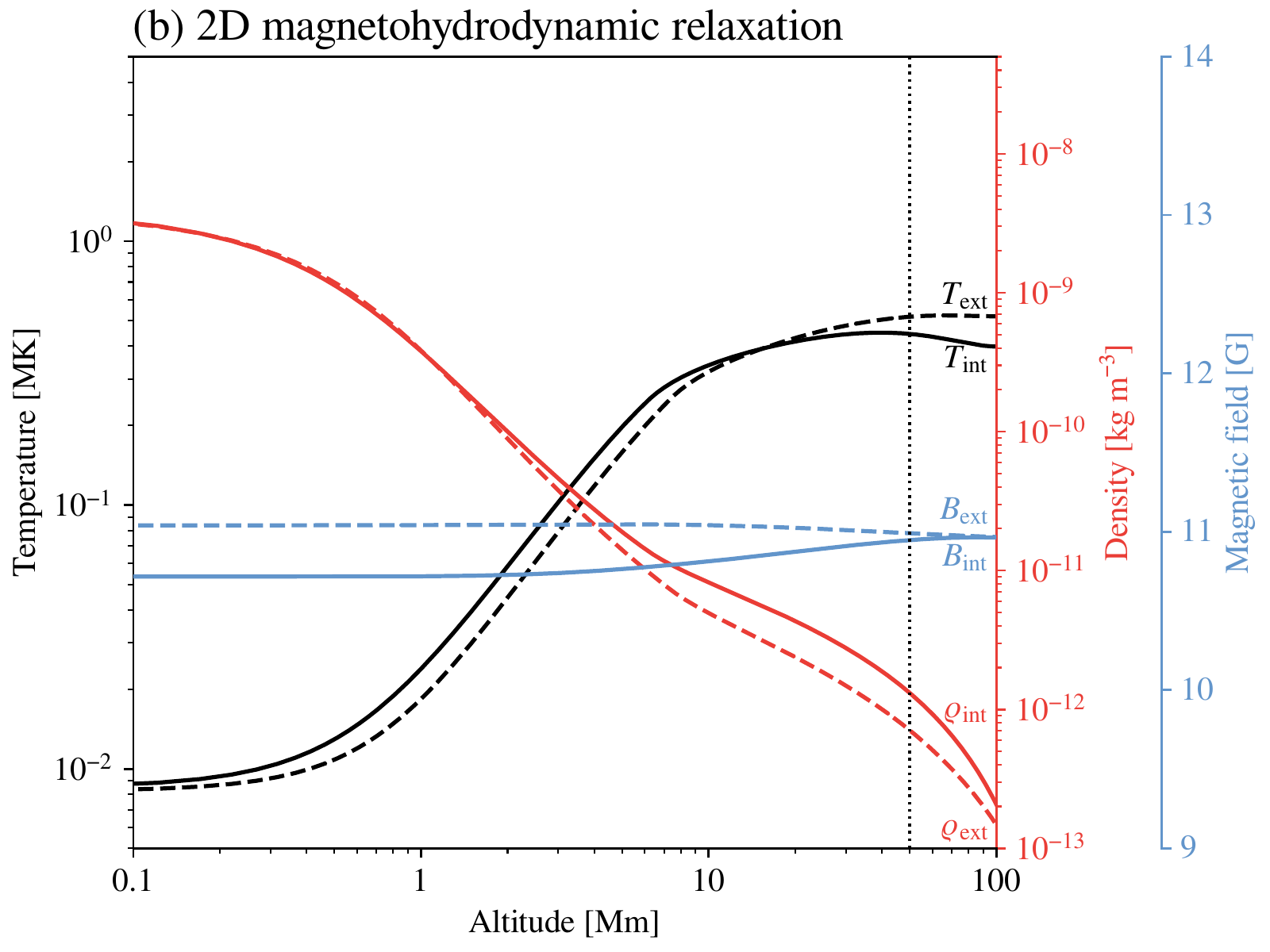}
\caption{
  Temperature (\textit{black}),
  density (\textit{red}),
  and magnetic field magnitude (\textit{blue})
  profiles
  inside ($r = \SI{0}{Mm}$; \textit{solid lines})
  and outside ($r = \SI{8}{Mm}$; \textit{dashed lines})
  the flux tube.
  (a)~After solving the field-aligned hydrostatic equilibrium.
  (b)~After the 2D magnetohydrodynamic relaxation.
  }
\label{fig:relax_profiles}
\end{figure*}

The simulation was initialized with a uniform vertical magnetic field of magnitude $B_0 = \SI{42}{G}$.
Along the tube, we imposed the following temperature profile, derived from \citet{AschwandenSchrijver2002}:

\begin{equation}
T(x, y, z) =
\begin{cases}
\Tch & \text{if $z \leq \Delta_\mathrm{ch}$,} \\
\Tch + \left(\Tcor(x, y) - \Tch\right) \left( 1 - \left(\frac{L - z}{L - \Delta_\mathrm{ch}} \right)^2 \right)^{0.3} & \text{else,}
\end{cases}
\end{equation}
where $z$ is the altitude,
$L$ is the height of the computational domain,
$\Delta_\mathrm{ch} = \SI{4}{Mm}$ is thickness of the chromosphere,
and $\Tch = \SI{20000}{K}$ is the temperature in the chromosphere.
We defined the transverse temperature profile at the top of the domain, $\Tcor(x, y)$, as:
\begin{equation}
\Tcor(x, y) = \Tcorext + (\Tcorint - \Tcorext) \zeta(x, y),
\end{equation}
where $\Tcorint = \SI{1.2}{MK}$ is the temperature inside the tube, and $\Tcorext = \SI{3.6}{MK}$ is the temperature outside the tube.
The shape of the profile was set by $\zeta(x, y)$:
\begin{equation}
\zeta(x, y) =
\frac{1}{2}
\left[
  1 -
  \tanh
  \left(
    \left(
      \sqrt{x^2 + y^2} / R - 1
    \right)
    b
  \right)
\right],
\end{equation}
where $R = \SI{1}{Mm}$ is the tube radius, and $b = 5$ is a dimensionless number setting the width of the inhomogeneous layer between the interior and exterior of the tube ($l \approx 6R/b$).
$\zeta(x, y)$ is close to \num{1} inside the tube, and to \num{0} outside.

We also set the density at the bottom of the chromosphere ($z = 0$) to:
\begin{equation}
\rhoch(x, y, z = 0) = \rhochext + (\rhochint - \rhochext) \zeta(x, y),
\end{equation}
where $\rhochint = \SI{3.51e-8}{kg.m^{-3}}$ is the density inside the tube, and $\rhochext = \SI{1.17e-8}{kg.m^{-3}}$ is the density outside.
We then integrated the field-aligned hydrostatic equilibrium equation numerically using a Crank-Nicholson scheme.
The profiles of the imposed temperature and of the density resulting from the integration are shown in \autoref{fig:relax_profiles}~(a).
The temperature contrast (interior temperature divided by exterior temperature) is \num{1} in the chromosphere, and decreases to \nicefrac{1}{3} in the corona.
The density contrast is \num{3} in the chromosphere, increases to around \num{7} in the transition region, and decreases again to about \num{4} in the upper corona.
The pressure contrast is \num{3} in the chromosphere, and slowly decreases to reach \num{1.2} in the upper corona.

However, this initial state is not in magnetohydrostatic (MHS) equilibrium, because the pressure varies across the flux tube, while the magnetic field does not.
To fix this, we let the tube relax by running a 2D magnetohydrodynamic simulation (\autoref{sec:model:relax_2D}).
We then used this relaxed state to initialize the 3D simulation of kink waves (\autoref{sec:model:driven_3D}).

\subsection{Flux tube relaxation (2D)}
\label{sec:model:relax_2D}

In order to obtain a flux tube in MHS equilibrium, we first run a 2D simulation, initialized with the initial state described in \autoref{sec:model:initial_cond}.
The MHD equations were solved in a longitudinal plane at $y = 0$ (see \autoref{fig:domain_sketch}), with $x \in [0, 8.56]~\si{Mm}$, and $z \in [0, 100]~\si{Mm}$.
We used a uniform grid of $64 \times 2048$ cells with a size of $\SI{134}{km}\times\SI{49}{km}$.
The resolution along $z$ is higher than in the 3D runs in order to resolve the sharper gradients in the transition region (see \autoref{fig:relax_profiles}).
We verified that a resolution of \SI{40}{km} in the $x$ direction yielded the same results, by running a separate 2D simulation followed by a 3D driven simulation ($P_0 = \SI{200
}{s}$), and verifying that the cut-off altitude and comparison to the analytical formulas (\autoref{sec:discussion}) were not significantly modified.

We let the system evolve for \SI{47}{ks}, during which the velocity rewrite parameter $\alpha_v$ varied as described in Eq. (\ref{eq:vrw_av2D}).
As a result of the relaxation, periodic longitudinal flows with a velocity of about \SI{15}{\kilo\meter\per\second} develop along the tube.
They are damped during the later stages of the simulation, as the velocity rewrite layer is gradually introduced.
At the end of the relaxation run, residual velocities are lower than \SI{0.5}{\kilo\meter\per\second} everywhere in the domain.
The resulting temperature, density, and magnetic field profiles are shown on \autoref{fig:relax_profiles}~(b).
Compared to the initial state (\autoref{fig:relax_profiles}~a), the transition region is significantly broadened, with a thickness of about \SI{7}{Mm}.
This is the direct result of the modified thermal conductivity used in this setup, and allows for a coarser resolution along the loop in the 3D simulations.
In addition, the temperature and density decrease, both inside and outside the tube.
Overall, the density contrast  ($\rho_\mathrm{int}/\rho_\mathrm{ext}$) decreases: it reaches \num{1} in the chromosphere, \num{1.2} in the transition region, and \num{1.8} in the corona.
The temperature contrast also changes to about \num{1.3} in the transition, and about \num{0.8} in the corona.
Finally, the magnetic field amplitude contrast remains very close to \num{1} everywhere in the domain (\num{0.97} in the chromosphere and \num{1} in the corona), with a magnitude of about \SI{11}{G} everywhere in the domain.
Compared to the initial uniform magnetic field, the magnitude is divided by about four, while the contrast remains close to \num{1}.
The final temperature and density profile significantly differ from the initial conditions of 2D relaxation run.
However, this is not an issue, as the goal of this study is to investigate how the analytical formulas we consider \citep{Spruit1981, LopinNagorny2017oct, SnowEtAl2017} predict the cut-off frequency for a given temperature and density profile.
By using the relaxed profiles as an input to these analytical formulas, we obtained predictions for the relaxed system.

This relaxed 2D simulation was then mapped onto the 3D domain through cylindrical symmetry.
We used a rotation about the line $x = 0$ (i.e. the centre of the loop), and a trilinear interpolation to project onto the 3D Cartesian grid.

\subsection{Kink waves propagation (3D)}
\label{sec:model:driven_3D}

\begin{figure*}
\includegraphics[width=\textwidth]{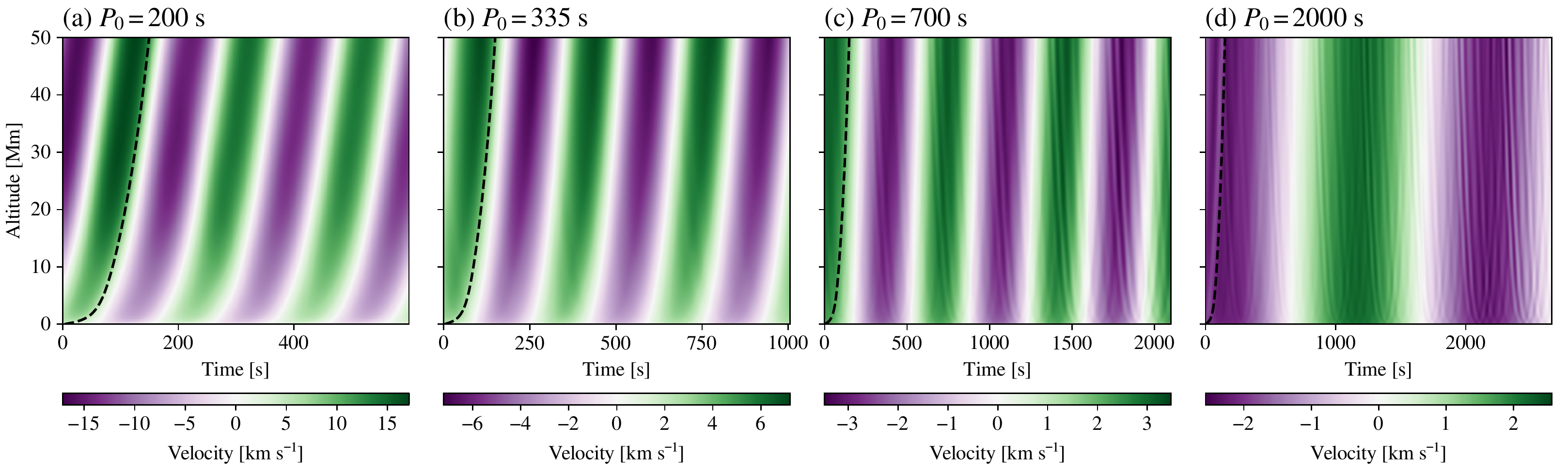}
\caption{Kink waves transverse velocity ($v_x$) at the loop centre ($x = y = 0$), as a function of altitude and time.
The velocity is shown for four 3D simulations with different driver periods $P_0$, after an initial settling time of $2 P_0$ (for $P_0 = \SI{200}{s}$, \SI{335}{s} and \SI{700}{s}), or $0.42 P_0$ (for $P_0 = \SI{2000}{s}$).
The dashed black lines represent a propagation at the kink speed (see Eq. (\ref{eq:ck})), and are independent of the driver period.
}
\label{fig:time_dist_v}
\end{figure*}

In order to simulate the propagation of kink waves from the chromosphere to the corona, we drove the 3D simulations with the monoperiodic, dipole-like, driver described in Eqs. (\ref{eq:driver_inside}) and (\ref{eq:driver_outside}).
We ran four simulations, with different driver periods $P_0$: \SI{200}{s}, \SI{335}{s}, \SI{700}{s}, and \SI{2000}{s}.
The propagating kink waves generated by the driver are absorbed by the velocity rewrite layer at the top of the domain, and are thus not reflected downwards.
The first three simulations were run for a duration of $5 P_0$.
The last simulation was run for $1.75 P_0$.
At the beginning of the simulations, the system goes through an initial transitory phase before the propagating kink wave is fully established (i.e. its amplitude does not change with time).
We waited for $2 P_0$ ($0.42 P_0$ for $P_0 = \SI{2000}{s}$) for the kink wave to enter a stable sinusoidal regime.
After this duration, we saved high-cadence snapshots at the centre of the loop (line $x = y = 0$).
For all further analysis, we used the snapshots saved after the transitory phase.
The transverse velocity $v_x$ at the loop centre is shown in \autoref{fig:time_dist_v}. 
As can be seen on this figure, the amplitude of the kink wave decreases as the period increases.
For the two longer driver periods (\num{700} and \SI{2000}{s}), the amplitude of the kink wave is small enough for some perturbations to become visible.
They travel at the Alfvén speed, and appear to be triggered by the flows remaining after the relaxation (see \autoref{sec:model:relax_2D}).
These perturbations have amplitudes smaller than \SI{0.2}{\kilo\meter\per\second}, and should thus have no effect on the wave.

\section{Results: cut-off and tunnelling of transverse waves}
\label{sec:res}

In order to determine whether the kink waves driven in the 3D simulations are experiencing a cut-off, we looked at the evolution of the velocity amplitude (\autoref{sec:res:vel_ampl}), as well as the phase speed (\autoref{sec:res:phase_speed}) as a function of altitude.
The analysis of these profiles allows us to establish that the transverse waves are subject to a low-frequency cut-off in the transition region.

\subsection{Wave amplitude increases with frequency}
\label{sec:res:vel_ampl}

\begin{figure}
\includegraphics[width=\columnwidth]{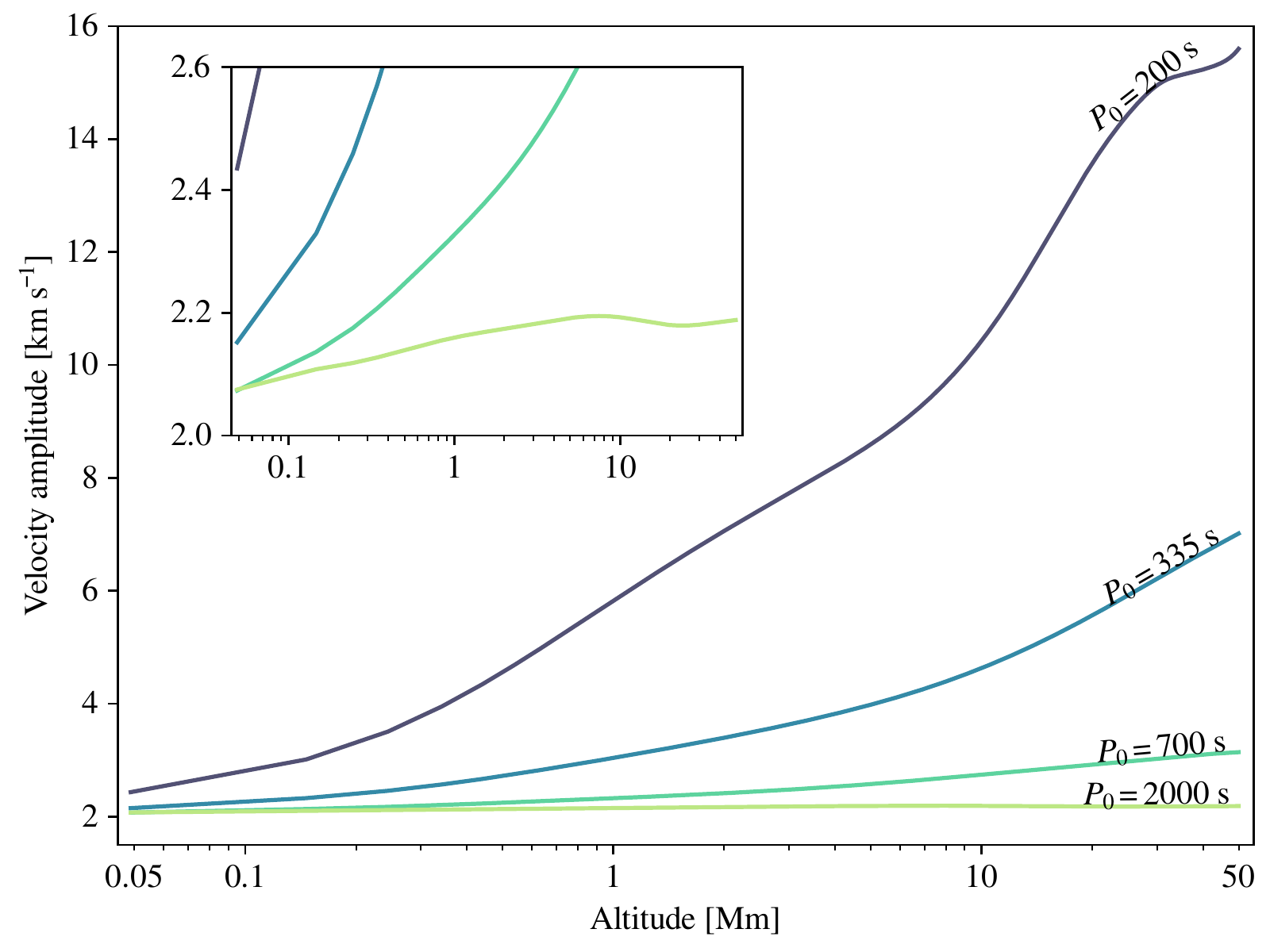}
\caption{Velocity amplitude of kink waves, as a function of altitude.
The velocity is shown for four different driver periods ($P_0$).
The inset has the same axes as the main figure, with a zoom-in on the vertical axis.}
\label{fig:ampl_alt}
\end{figure}

In order to compute the velocity amplitude of the kink wave, we fitted the function $A_x(z) \sin\left( \omega(z) t + \phi(z) \right)$ to the transverse velocity $v_x(z, t)$, at each altitude ($z$).
$A_x(z)$ is the velocity amplitude, $\omega(z)$ is the kink wave frequency, and $\phi(z)$ is the phase.
The frequency varies by less than \SI{1}{\percent} with altitude, confirming theoretical understanding.
The velocity amplitude is shown in \autoref{fig:ampl_alt}.
In all simulations, the wave amplitude increases with altitude, because of the density decreases with altitude and energy conservation.
Across simulations, the amplitude at a given altitude increases with the frequency of the wave.
This means that kink waves with higher frequencies propagate better from the chromosphere to the corona.
This would be consistent with the low-frequency cut-off predicted by analytical models (see \autoref{sec:introduction}).

\subsection{Evanescent waves in the transition region}
\label{sec:res:phase_speed}

To determine the altitude at which the waves are cut-off, we compared their phase speed $v_p(z)$ to the kink speed of the flux tube $c_k(z)$.
The inverse phase speed is equivalent to the phase difference $\Delta\phi(z)$ between two altitudes separated by $\Delta z$: $1/v_p(z) = \Delta\phi(z)/(\omega \Delta z)$.
The phase difference has been successfully used to determine the cut-off frequency of acoustic and slow-magnetosonic waves in observations \citep{CentenoEtAl2006, FelipeEtAl2010, KrishnaPrasadEtAl2017, FelipeEtAl2018}, and in simulations \citep{FelipeSangeetha2020}.
In these articles, the authors determine the phase speed for a wide range of frequencies, but at a limited number of altitude positions.
In the present study however, we could only examine four frequencies, because of the high computational cost of a simulation.
However, we computed the phase difference at all altitudes of the simulation domain.
This allows us to determine the altitude at which the wave is cut-off.

The phase speed at a given altitude $z$ was computed from the transverse velocity in the cells above and below, that is $v_x(t, z+\Delta z / 2)$ and $v_x(t, z-\Delta z / 2)$, where $\Delta z = \SI{98}{km}$ is the cell size.
We apodized these velocity time series with a Hann window, and computed the cross-correlation $C(\tau, z) = v_x(t, z+\Delta z / 2) \star v_x(t, z-\Delta z / 2)$.
We then determined the time delay $\Delta\tau(z)$, by finding the maximum of $C(\tau, z)$.
To that end, we fitted the function $A + B \cos\left(\omega(\tau - \Delta\tau) / \delta\right)$ to $C(\tau, z)$, with $\tau \in [-P_0/4, +P_0/4]$.
Finally, the phase difference was given by $\Delta\phi(z) = \omega \Delta\tau(z)$, and the inverse phase speed by $1/v_p(z) = \Delta\tau(z) / \Delta z$.
The inverse phase speed is shown on \autoref{fig:phase_alt}, alongside the inverse kink speed for the simulated flux tube.
The kink speed $c_k$ is calculated using:
\begin{equation}
\label{eq:ck}
c_k^2(z) = \frac{\rho_i(z) v_{A\,i}^2(z) + \rho_e(z) v_{A\,e}^2(z)}{\rho_i(z) + \rho_e(z)},
\end{equation}
where $\rho(z)$ is the density, $v_A(z) = B(z) / \sqrt{\mu_0 \rho(z)}$ is the Alfvén speed, $B(z)$ is the magnetic field amplitude, and $\mu_0$ is the magnetic permittivity of vacuum.
The indices $i$ and $e$ correspond, respectively, to internal and external quantities relatively to the flux tube, and are taken at $x = 0$ and $x = \SI{8}{Mm}$.

In simulations with short driver periods, the inverse phase speed is somewhat smaller than the inverse kink speed in the chromosphere and transition region ($v_p/c_k \approx 2$ for $P_0 = \SI{200}{s}$, and $5$ for $P_0 = \SI{335}{s}$), and equals the inverse kink speed in the corona.
On the other hand, in simulations with longer periods, the inverse phase speeds are much lower than the inverse kink speed below a given altitude.
For $P_0 = \SI{700}{s}$, $1/v_p$ is about \num{250} times smaller than $1/c_k$ below $z = \SI{1}{Mm}$. For $P_0 = \SI{2000}{s}$, a similar drop occurs below $z = \SI{20}{Mm}$.

For a propagating kink wave, the inverse phase speed is expected to be equal to the inverse kink speed.
Conversely, standing and evanescent (i.e. cut-off) waves have inverse phase speeds smaller than the inverse kink speed.
Thus, the decreased inverse phase speed for higher periods indicates that the waves are cut-off in at least some regions.

To distinguish between the standing and evanescent cases, we have also looked at the wave amplitude (\autoref{fig:ampl_alt}).
In the absence of vertical stratification, the amplitude of evanescent waves decreases with altitude.
However, in a stratified atmosphere (our  case), the amplitude increases with altitude because of the density decrease, even for evanescent waves.
On \autoref{fig:ampl_alt}, the amplitude of waves with longer periods (for which $1/v_p \ll 1/c_k$) increases less with altitude compared to waves with shorter periods (for which $1/v_p \lesssim 1/c_k$).
We thus conclude that the waves with longer periods are evanescent in parts of the low atmosphere, where their inverse phase speed is much lower than the inverse kink speed.
This means that these long-period waves are cut-off in the transition region.

\subsection{Wave tunnelling at higher frequencies}
\label{sec:res:tunnelling}

\begin{figure}
\includegraphics[width=\columnwidth]{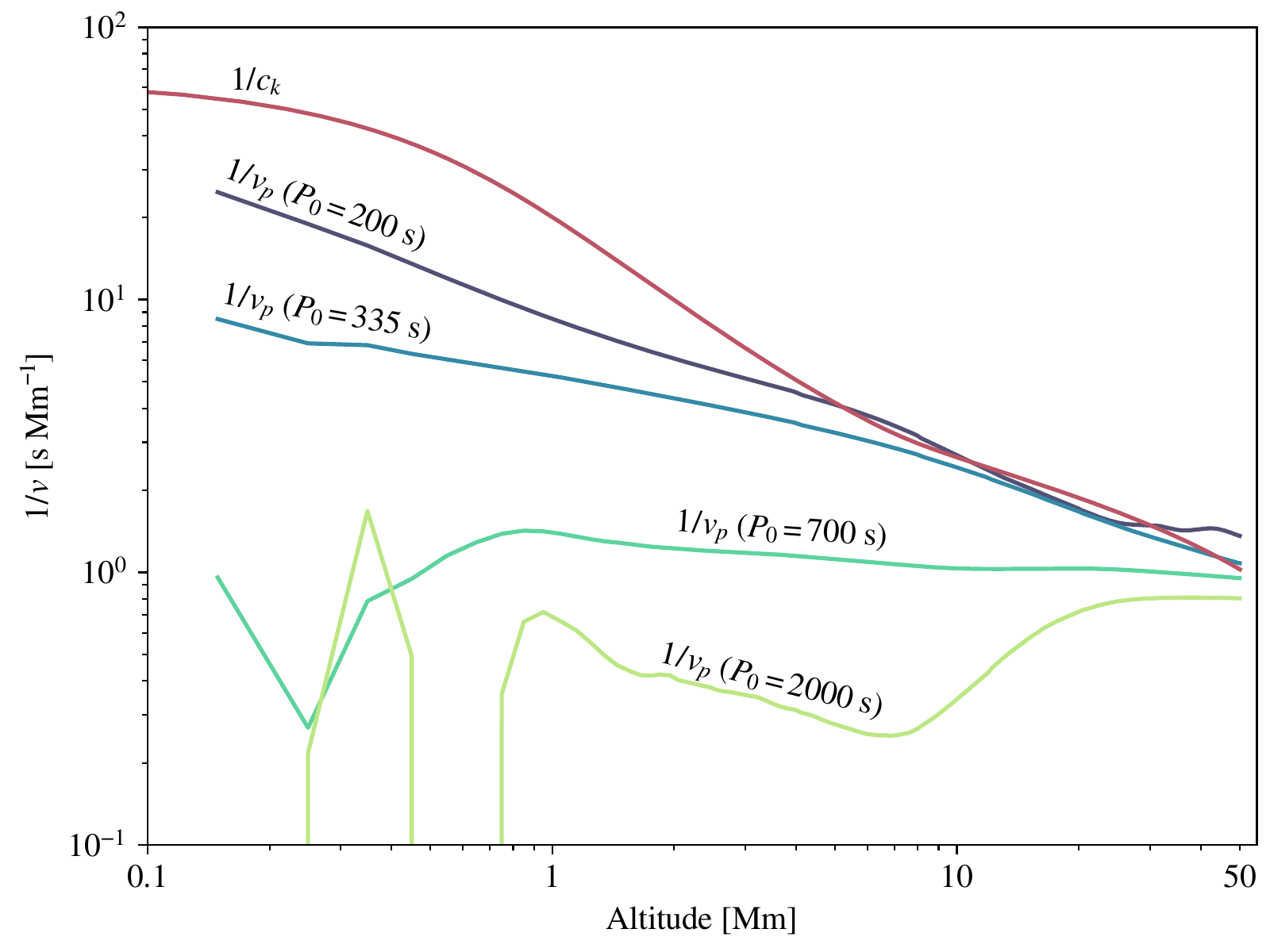}
\caption{Inverse phase speed of the kink wave ($1 / v_p$), and inverse kink speed of the flux tube ($1 / c_k$), as a function of altitude.
The phase speed is given for four different driver periods ($P_0$).
}
\label{fig:phase_alt}
\end{figure}

Waves with shorter periods ($P_0 = \num{200}$ and \SI{335}{s}) also show signs of cut-off at low altitudes.
Below $z = \SI{3}{Mm}$, the inverse phase speed $1/v_p$ is lower than the inverse kink speed $1/c_k$ (\autoref{fig:phase_alt}), and the amplitude increase with altitude is smaller for $P_0 = \SI{335}{s}$ than for $P_0 = \SI{200}{s}$ (\autoref{fig:ampl_alt}).
However, this cut-off is significantly weaker than in the long-period case.
This is explained by the fact that the cut-off region (where $1/v_p < 1/c_k$) is narrower for short periods ($\sim \SI{1}{Mm}$) than for long periods ($\sim \SI{10}{Mm}$).
As a result, short-period waves can tunnel through the cut-off region, and propagate into the corona.
Furthermore, the weak attenuation in the cut-off region ($1/v_p \lesssim 1/c_k$) results further reduces the effect of the cut-off.

\section{Discussion: comparison to analytical formulas}
\label{sec:discussion}

\begin{figure}
\includegraphics[width=\columnwidth]{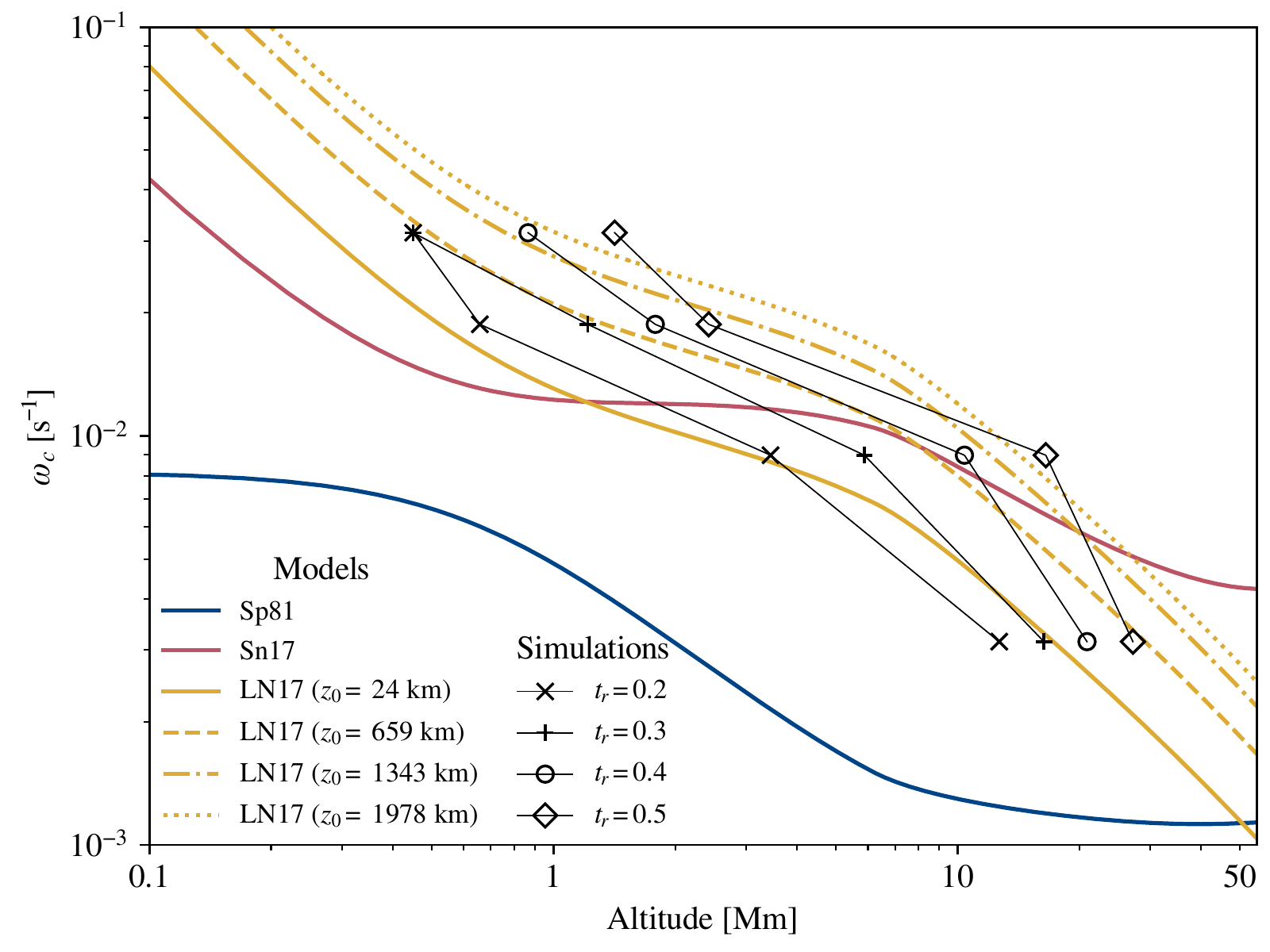}
\caption{Kink wave cut-off frequency as a function of altitude, from analytical models (left column of the legend), and from our numerical simulations (right column of the legend).
We show the analytical predictions of \citet[][SP81]{Spruit1981}, \citet[][Sn17]{SnowEtAl2017}, and of \citet[][LN17]{LopinNagorny2017oct} (\textit{coloured lines}).
For the last model, we computed the cut-off frequency for different values of $z_0$, the “base of the atmosphere”.
We show the cut-off altitude ($z_c$) for the four simulations that we ran with different driver frequencies (\textit{black markers}).
The cut-off altitudes are computed with different thresholds $t_r$, indicated on the legend and described in the text.
}
\label{fig:cutoff_frequency}
\end{figure}

In order to compare our simulations to the analytical models, we quantified the cut-off frequency as a function of altitude.
We define $z_c$, the altitude at which $c_k / v_p$ goes above a given threshold $t_r$.
This corresponds to the altitude where the wave leaves the cut-off regime and enters the propagating regime.
That is, the cut-off altitude.
We computed $z_c$ for four values of $t_r$ between \num{0.2} and \num{0.5}.
Considering the four simulations with different driver frequencies $\omega$, we obtained the cut-off altitude as a function of the frequency, $z_c(\omega)$.
We compare this to the cut-off frequency as a function of altitude, $\omega_c(z)$, predicted by the analytical models presented in \autoref{sec:introduction}.

On \autoref{fig:cutoff_frequency}, we show the cut-off frequency and altitude computed in our simulations, for different values of $t_r$ (\textit{black} points).
On the same figure, we show the predictions of the analytical formulas of \citet[][Eq.~(\ref{eq:wcSp81})]{Spruit1981}, \citet[][Eq.~(\ref{eq:wcLN17})]{LopinNagorny2017oct}, and \citet[][Eq.~(\ref{eq:wcSn17})]{SnowEtAl2017} (\textit{coloured lines}), computed for the temperature and density profiles used in our simulations.
We implement the formula of \citet{LopinNagorny2017oct} for different values of $z_0$, defined by the authors as “the base of the atmosphere”, with no further details.
Because this quantity is not accurately defined, we used four values of $z_0$ in the range of \SI{24}{km} (bottom cell of our simulation domain), to \SI{1978}{km}.
This loosely defined parameter broadens the range for the cut-off frequencies predicted by this formula.
While the match is rather loose, the cut-off altitude $z_c(\omega)$ measured in our simulations matches the overall variation the cut-off frequency $\omega_c(z)$ predicted by the \citet{LopinNagorny2017oct} formula.
In particular, the shape of the profiles are in good agreement.
On the contrary, the \citet{SnowEtAl2017} model correctly predicts the cut-off frequency only in the lower transition region, but fails to do so in the upper transition region and corona.
In particular, their model predicts a slower decrease of the cut-off frequency above \SI{20}{Mm}, while the simulations and the \citet{LopinNagorny2017oct} show a continued decrease.
Finally, the \citet{Spruit1981} predictions are off by almost an order of magnitude at all altitudes.
Thus, the formula of \citet{LopinNagorny2017oct} best predicts the cut-off frequency of transverse waves at different altitudes.

While the broadened transition region in our simulations could affect the altitude-dependence of the cut-off frequency, this should have little impact on the validation of the analytical formulas.
Indeed, these formulas include the atmospheric stratification through altitude-dependent profiles of either the pressure scale height or the Alfvén speed (see \autoref{sec:introduction}).
Because they make no hypothesis on these profiles, they should be valid regardless of the atmosphere considered.
As such, the agreement with the simulations should not depend on the broadening of the transition region, provided the appropriate profile is fed into the formulas.
After validating the \citet{LopinNagorny2017oct} formula by comparing it to our simulations, it should be applicable to other stratification profiles.

We note that while analytical formulas can predict the kink cut-off frequency, this is not sufficient to know whether a kink wave with a given frequency will propagate into the corona.
To that end, the thickness of the cut-off region and the strength of the attenuation have to be taken into account.
As shown by our simulations, kink waves with higher frequencies ($\geq\SI{3}{mHz}$) can propagate into the corona by tunnelling through a region where they are cut-off (\autoref{sec:res:tunnelling}).
Furthermore, these waves only experience a weak attenuation, because their frequency is close to the cut-off frequency.
In fact, the cut-off frequency does not constitute a clear-cut boundary between oscillatory and non-oscillatory solutions.
This was also reported for sound waves by \citet{FelipeSangeetha2020}.
Although the question of whether a solution is oscillating is well-defined mathematically, this is not straightforward to translate into a single cut-off frequency \citep{SchmitzFleck1998}.
For this reason, there exist several canonical definitions for cut-off frequencies, set within the continuous variation between the oscillating and non-oscillating regimes (see e.g. \citealp{SchmitzFleck1998} for sound waves in the solar atmosphere).
As a result, cut-off frequencies are bound to be mere indications, rather than strong constraints, on the physical behaviour of a wave \citep{ChaeLitvinenko2018}.

\section{Conclusions}
\label{sec:conclusions}

Transverse waves are a candidate mechanism for heating the solar corona.
However, several analytical models predicted that they are cut-off in the transition region.
In order to assess whether transverse waves can indeed heat the corona, it is thus crucial to determine whether they can propagate through the transition region.
To that end, we have simulated the propagation of transverse kink waves in an open magnetic flux tube, embedded in an atmosphere extending from the chromosphere to the corona.
We found that transverse waves are indeed cut-off in the lower solar atmosphere.
However, only waves with low frequencies ($\nu \lesssim \SI{2}{mHz}$) are significantly affected.
At higher frequencies, the cut-off occurs in a very thin layer ($\sim \SI{1}{Mm}$), and results in a weak attenuation. 
In this case, waves can tunnel through the cut-off layer, experiencing little to no amplitude attenuation.
This means that transverse waves with high frequencies are able to transport energy from the chromosphere to the corona, where it can be dissipated and result in heating.

Furthermore, we compared our simulations to several analytical models that predict the cut-off frequency of transverse waves.
We conclude that the formula proposed by \citet{LopinNagorny2017oct} gives the best prediction.
While our simulations use a broadened transition, we expect it to have little impact on the validation of analytical formulas.
As such, the formula by \citet{LopinNagorny2017oct} should be able to predict the cut-off frequency for any atmospheric stratification profile.
We note that while the cut-off frequency is a good first indicator of whether a wave can propagate into the corona, it cannot alone predict the whole behaviour of the wave.
In particular, waves with frequencies just below the cut-off frequency (that should thus be cut-off) can still reach the corona, thanks to a combination of tunnelling, and weak attenuation.

\begin{acknowledgements}
This project has received funding from the European Research Council (ERC) under the European Union’s Horizon 2020 research and innovation program (grant agreement No. 724326).
GP was supported by a CNES postdoctoral allocation.
TVD was supported by the European Research Council (ERC) under the European Union's Horizon 2020 research and innovation programme (grant agreement No 724326) and the C1 grant TRACEspace of Internal Funds KU Leuven. 
K.K. recognises support from a postdoctoral mandate from KU Leuven Internal Funds (PDM/2019), from a UK Science and Technology Facilities Council (STFC) grant ST/T000384/1, and from a FWO (Fonds voor Wetenschappelijk Onderzoek – Vlaanderen) postdoctoral fellowship (1273221N). The results received support from the FWO senior research project with number G088021N.
\textit{Software:}
Astropy \citep{Astropy2013, Astropy2018},
\end{acknowledgements}

\end{document}